\documentclass{article}

\usepackage{PRIMEarxiv}

\usepackage[utf8]{inputenc} 
\usepackage[T1]{fontenc}    
\usepackage{hyperref}       
\usepackage{url}            
\usepackage{booktabs}       
\usepackage{multirow}       
\usepackage{amsfonts}       
\usepackage{nicefrac}       
\usepackage{microtype}      
\usepackage{fancyhdr}       
\usepackage{graphicx}       
\usepackage{algorithm,algorithmic}  
\usepackage{amsmath,amssymb,amsfonts}  
\usepackage[style=ext-authoryear]{biblatex}
\graphicspath{{figures/}}     

\title{ConvDTW-ACS: Audio Segmentation for Track Type Detection During Car Manufacturing}

\author{
\'Alvaro L\'opez-Chilet \\
\textit{PRHLT Research Center} \\
\textit{Universitat Polit\`ecnica de Val\`encia} \\
Valencia, Spain \\
allochi@upv.es \\
\and
Zhaoyi Liu \\
\textit{imec-Distrinet} \\
\textit{KU Leuven} \\
Leuven, Belgium \\
zhaoyi.liu@kuleuven.be \\
\and
Jon Ander G\'omez \\
\textit{PRHLT Research Center} \\
\textit{Universitat Polit\`ecnica de Val\`encia} \\
Valencia, Spain \\
jon@upv.es \\
\and
Carlos Alvarez \\
\textit{Launch Department} \\
\textit{Ford Motor Company} \\
Valencia, Spain \\
calvare5@ford.com \\
\and
Marivi Alonso Ortiz \\
\textit{Quality Department} \\
\textit{Ford Motor Company} \\
Valencia, Spain \\
malonsoo@ford.com \\
\and
Andres Orejuela Mesa \\
\textit{Quality Department} \\
\textit{Ford Motor Company} \\
Valencia, Spain \\
aorejuel@ford.com \\
\and
David Newton \\
\textit{Office of the CIO - Security \& Strategy} \\
\textit{Ford Motor Company} \\
Cologne, Germany \\
dnewton@ford.com \\
\and
Friedrich Wolf-Monheim \\
\textit{Ford Research \& Advanced Engineering} \\
\textit{Ford Motor Company}\\
Aachen, Germany \\
fwolf5@ford.com \\
\and
Sam Michiels \\
\textit{imec-Distrinet} \\
\textit{KU Leuven} \\
Leuven, Belgium \\
sam.michiels@kuleuven.be \\
\and
Danny Hughes \\
\textit{imec-Distrinet} \\
\textit{KU Leuven} \\
Leuven, Belgium \\
danny.hughes@kuleuven.be \\
}

\begin{document}
\maketitle

\begin{abstract}
This paper proposes a method for Acoustic Constrained Segmentation (ACS) in audio recordings of vehicles driven through a production test track, delimiting the boundaries of surface types in the track. ACS is a variant of classical acoustic segmentation where the sequence of labels is known, contiguous and invariable, which is especially useful in this work as the test track has a standard configuration of surface types. The proposed ConvDTW-ACS method utilizes a Convolutional Neural Network for classifying overlapping image chunks extracted from the full audio spectrogram. Then, our custom Dynamic Time Warping algorithm aligns the sequence of predicted probabilities to the sequence of surface types in the track, from which timestamps of the surface type boundaries can be extracted. The method was evaluated on a real-world dataset collected from the Ford Manufacturing Plant in Valencia (Spain), achieving a mean error of $166$ milliseconds when delimiting, within the audio, the boundaries of the surfaces in the track. The results demonstrate the effectiveness of the proposed method in accurately segmenting different surface types, which could enable the development of more specialized AI systems to improve the quality inspection process.
\end{abstract}


\section{Introduction} \label{sec:1_introduction}

Artificial Intelligence (AI) is a vital tool in modern manufacturing, particularly in Industry 4.0. By leveraging technologies like 5G~(\cite{industry_5g2022}), interconnected devices facilitate real-time data collection, analysis, and predictions. 

One significant application of AI in vehicle manufacturing is process automation, leading to increased efficiency and cost reduction. A critical research area focuses on detecting and analyzing faults in quality control~(\cite{industry_fault_detection2022}). This study centres on the quality inspection phase at the Ford Valencia Manufacturing Plant, aiming to develop a system that identifies vehicles deviating from established quality standards. A sound evaluation test is conducted post-assembly on a standardized track with various surfaces simulating different driving conditions. Skilled personnel listen for abnormal sounds indicating manufacturing defects. Automating this test requires a robust AI model to achieve high performance. Developing a segmentation model to categorize audio data by surface type enables more effective operation of specialized fault detection systems.

In this work, we present ConvDTW-ACS, a method to automatically detect the boundaries between different track surfaces given the audio of a test track run. This is an Acoustic Constrained Segmentation (ACS) task where the sequence of segments is known, contiguous and invariant, as all the recordings are collected in the same track. Therefore, the proposed method exploits this constraint to generate a better segmentation. First, a spectrogram is extracted from the whole audio. Second, a sequence of overlapping patches (\textit{chunks}) are extracted along the time axis of the spectrogram. Then, a Convolutional Neural Network (CNN) classifies the chunks into their corresponding surface label. Finally, a constrained variant of the Dynamic Time Warping algorithm for ACS (ACS-DTW) is used to align the model predictions to the sequence of surfaces in the track, correcting errors and improving the segmentation.

The main contributions of this work are:

\begin{itemize}
    \item A method called ConvDTW-ACS for ACS. It is designed to take into account the constraints of the task to create a more precise segmentation.
    \item The model is evaluated using real-world data obtained during production at the Ford Manufacuring Plant in Valencia (Spain).
    \item A review of the hyperparameter optimization results is presented to analyze the impact on the performance of different data preprocessing configurations, model sizes and training techniques.
\end{itemize}

The rest of the paper is organized as follows: Section~\ref{sec:2_related_work} introduces some of the related techniques in the state-of-the-art. Section~\ref{sec:3_proposed_method} gives a full description of the problem and the proposed method. Section~\ref{sec:4_experimental_setup} describes the dataset, metrics and tools used for experimentation. Section~\ref{sec:5_results} shows the results obtained and the exploration of hyperparameters carried out. Finally, section~\ref{sec:6_conclusions} presents our conclusions.
\section{Related Work} \label{sec:2_related_work}

In this section, we discuss some of the related tasks that can be found in literature, highlighting their similarities with ACS.

\textbf{Sound Event Detection (SED)} is a task that involves identifying a class of sound events and estimating the time position (i.e. start and end) of each occurrence of that class in an audio recording. Applications of SED include detecting noise sources in urban areas~(\cite{bello2019sonyc}), sound-based home monitoring~(\cite{martin2023strong}), and detecting anomalous machine sounds from manufacturing systems~(\cite{chen2023sw}).

The conventional approach to building automatic SED systems, especially automatic sound annotation, is to train a machine learning model with labelled data. Traditional methods include Gaussian Mixture Models (GMM)~(\cite{vuegen2013mfcc}), Decision Trees~(\cite{Decision2009}), Hidden Markov Models (HMM)~(\cite{heittola2013context}), and Support Vector Machines (SVM)~(\cite{portelo2009non}). Recent SED systems often use Deep Neural Networks (DNN)~(\cite{zhao2022seed, ronchini2022benchmark}), which are currently the state-of-the-art. In~\cite{yoho_segmentation_and_sed2022}, the authors proposed a method that divides the spectrogram into chunks of 0.303 seconds and applies spectrogram data augmentation (\textit{SpecAument}). They employ a Convolutional RNN model and perform a postprocessing step to remove spurious events. \cite{martin2023training} uses soft labels for SED and explores their benefits in capturing label distribution details and improving system performance in detecting missed sounds compared to hard labels.

In addition to SED, there are other related tasks, such as Audio Segmentation and Speaker Diarization. \textbf{Audio Segmentation} involves segmenting audio data into meaningful segments, often combined with other tasks. Deep Neural Networks (DNNs) are prevalent in state-of-the-art works, but end-to-end models are not commonly used; instead, postprocessing modules are frequently employed~(\cite{audio_segmentation_review2022}). In \cite{RNN_multiclass_segmentation_2020}, the authors segment classes like speech, music, and noise using Mel spectrograms along with chroma features and first/second-order derivatives of the features to include audio dynamics. They train a Conv1D model with Bidirectional LSTM and use an HMM for postprocessing, employing MixUp data augmentation. In \textbf{Speaker Diarization}, similar techniques are used. State-of-the-art approaches~(\cite{diarization_review2022}) typically employ DNNs for processing audio segments, followed by a post-processing step using methods such as the Viterbi algorithm~(\cite{diarization_viterbi_gmm}) or Bayesian Hidden Markov Models (BHMM)~(\cite{diarization_bhmm}) to obtain the final prediction. 

In the proposed ACS-DTW custom algorithm, step constraints are added to better align the model predicted probabilities with the known surface types order and their expected duration. These kind of restrictions have been previously applied in Speech Segmentation. In~\cite{speech_segmentation_dtw}, phonetic boundaries are detected using Acoustical Clustering-Dynamic Time Warping (AC-DTW), using the predicted probabilities of a GMM.
\section{Proposed Method} \label{sec:3_proposed_method}

This section presents first the problem description and then a complete revision of our proposed method.

\subsection{Problem Description}

In car manufacturing, after a vehicle is fully assembled it needs to pass several test phases to validate that the product meets the quality standards. Once the car exits the production line, fully assembled, it has to pass a drive test where different checks are performed. One of the main tests during this drive is to look for possible anomalous sounds that could be produced by a manufacturing defect. The test track where this final validation is performed has different kinds of surfaces. Each of these surfaces is designed to test specific driving conditions, from normal to more extreme situations. Therefore, anomalous sounds may appear in some surfaces and not in others.

The process timing is also very important when dealing with a complex and big production line that assembles hundreds of vehicles a day. Because of this, the inference time of the ACS system must be small, as it is going to be usually executed before other systems (e.g. sound anomaly detection).

The main difference between ACS and other typical segmentation tasks (e.g. music and speech detection~(\cite{audio_segmentation_review2022}) or speaker diarization~(\cite{diarization_review2022})) is that we know in advance which is the sequence of labels (surfaces) and that they are contiguous. In the end, the system should provide a list of boundaries separating the different surfaces of the track.

\subsection{Method Overview}

\begin{figure}[ht!]
\centerline{\includegraphics[width=0.5\columnwidth]{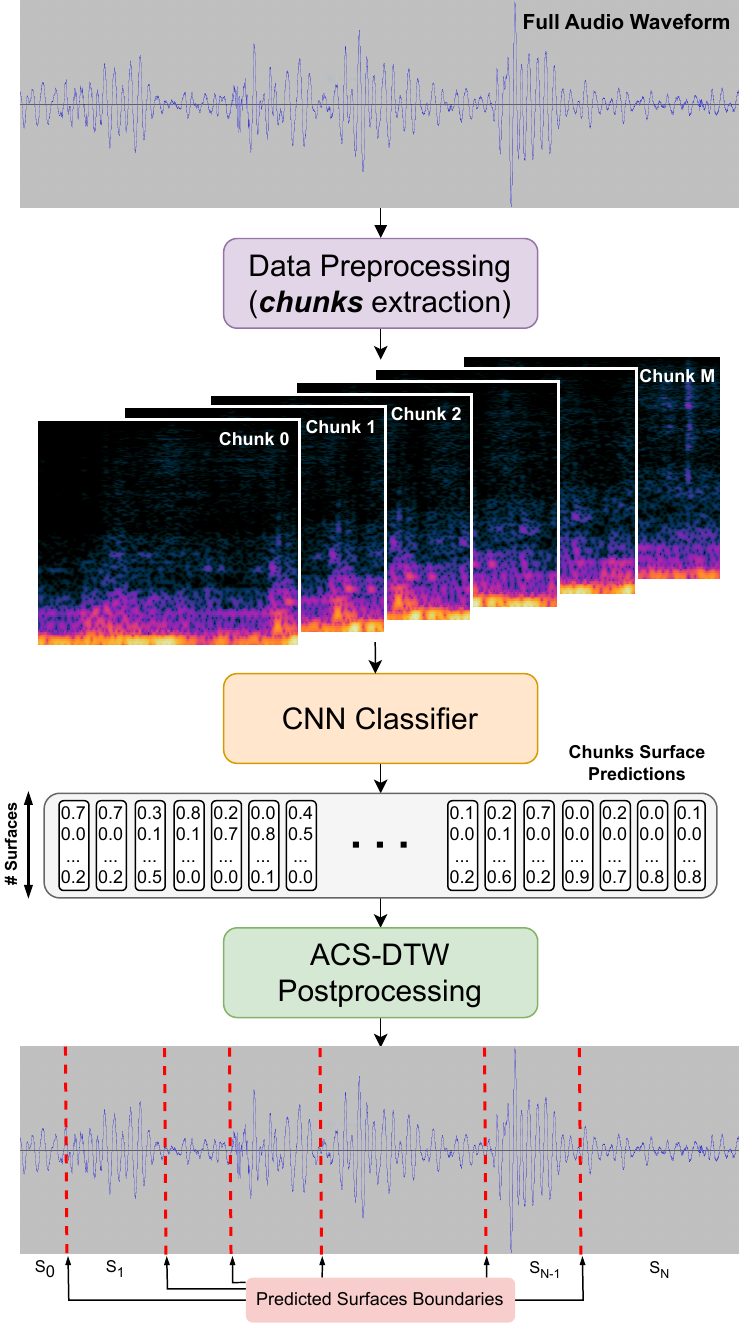}}
\caption{Diagram of the proposed ConvDTW-ACS method. First, the raw waveform is preprocessed to extract the spectrogram chunks. Then, a CNN model is used to classify each chunk among all the surfaces. Finally, our ACS-DTW algorithm is applied to align the model predictions to the order of the surfaces in the track. This produces the final segmentation composed by the boundary timestamps between surfaces.}
\label{fig:convdtw_acs}
\end{figure}

In this section we introduce the main blocks of ConvDTW-ACS to exploit the constraints of ACS. In order to predict the surface boundaries, the system proposed uses a Convolutional Neural Network (CNN) to perform audio classification over spectrogram segments, which we refer to as \textit{chunks}. For each chunk a surface label is assigned to train the classifier. Then, a postprocessing step using a variant of DTW is applied over the sequence of model predictions to get the final segmentation. In Figure~\ref{fig:convdtw_acs} a complete diagram of ConvDTW-ACS is shown.

To explain the method in more detail, it can be divided into three main blocks:

\begin{enumerate}
    \item \textbf{Data preprocessing:} Extracts a spectrogram from the raw waveform and generates fixed size chunks from it. The chunks are extracted along the time axis with overlap between them.
    \item \textbf{CNN classifier:} Performs the test track surface classification for each spectrogram chunk. This creates a sequence of probability vectors with a preliminary segmentation.
    \item \textbf{ACS-DTW Segmentation Postprocessing:} This block is the main component for ACS. Applies a custom variant of the Dynamic Time Warping (DTW) algorithm over the model predictions to correct and improve the segmentation. This optimization takes into account the order of the surfaces in the track to get the best alignment with the predictions. Therefore, chunk classification errors produced by the CNN model are corrected and the surface boundaries improved.
\end{enumerate}

In the following sections each pipeline block is explained in detail.

\subsection{Data Preprocessing}

The duration of one audio in the test track lasts about 2 minutes, including some out-of-track time where the car enters and leaves the track. This duration depends both on the speed of the driver and possible stops due to traffic or other production conditions. To deal with this long and variable audio samples, a chunk-based preprocessing is performed to get fixed size windows of data and process the whole signal.

\begin{figure}[ht!]
\centerline{\includegraphics[width=0.5\columnwidth]{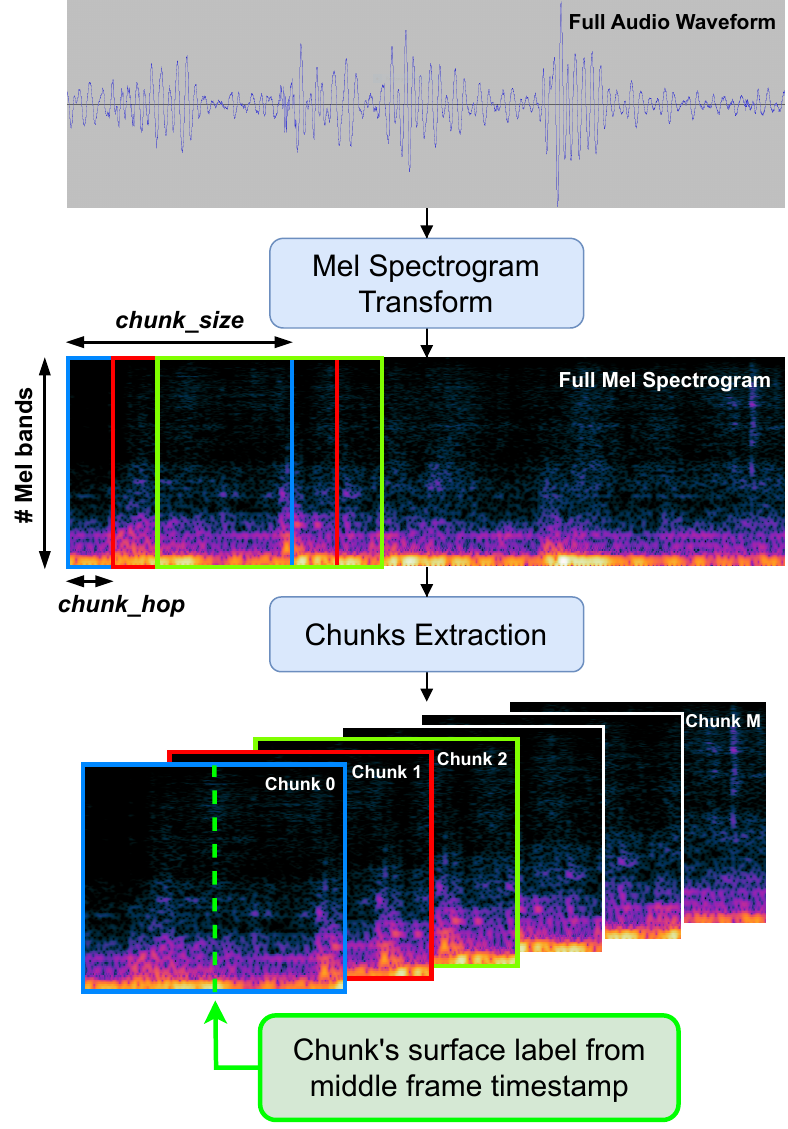}}
\caption{Diagram of the data preprocessing pipeline using the Mel spectrogram as feature extraction. It shows how the input samples (\textit{chunks}) for the CNN classifier are extracted from the raw audio waveform. The dimensions of each chunk image are $(channels=1, height=Mel\ bands, width=chunk\_size)$. For each chunk a surface label is assigned taking into account the middle frame of the chunk, generating the ground-truth for the CNN classifier.}
\label{fig:data_preprocessing}
\end{figure}

From the raw WAV file data to the final input data tensors feed to the model, our data loading pipeline performs the following steps (shown in Figure~\ref{fig:data_preprocessing}):

\begin{enumerate}
    \item \textbf{Waveform preprocessing:} First, the full waveform data is loaded. Then, in the experiments with waveform data augmentation, the following transformations from the \textit{audiomentations}~(\cite{audiomentations}) library were randomly applied: \textit{AirAbsorption}, \textit{AddGaussianSNR}, \textit{ClippingDistortion}. After the augmentations, the \textit{Normalize} transform is used to obtain always data in the same range $[-1, 1]$.
    
    \item \textbf{Wave to spectrogram:} It is well-stablished that transforming waveform data from the amplitude domain to the frequency domain can result in a better feature extraction~(\cite{diarization_review2022, audio_segmentation_review2022}). In our experiments, we tried three different spectrogram extractions from the \textit{torchaudio}~(\cite{torchaudio}) library: \textit{Spectrogram}, \textit{MelSpectrogram} and \textit{MFCC}.
    
    \item \textbf{Chunks extraction}: After getting the spectrogram of the whole audio, we extract chunks of fixed size to get the input samples for the model. The extraction is controlled by two hyperparameters: \textit{chunk\_size}, that determines the width of the chunk; and \textit{chunk\_hop}, that determines the displacement between chunks (see Figure~\ref{fig:data_preprocessing}). The \textit{chunk\_hop} controls the overlapping between chunks and determines the segmentation resolution. Where smaller values of it increase the resolution at the cost of increasing the number of chunks and the computational cost. Finally, to get the test track's surface label for each chunk, we take the label at the timestamp of the middle frame of the chunk, as more than one surface may appear in a chunk.  In the experiments with data augmentation, we also use \textit{MixUp} (\cite{mixup}) among the chunks of each training batch.
\end{enumerate}

\subsection{CNN Classifier} \label{subsec3:cnn_classifier}

Once all the $M$ chunks are extracted from the full audio sample, we perform a classification step to get $M$ predictions. All the chunks are fed into a CNN to classify among all the possible surfaces. The CNN architecture used is the \textit{ResNet}~(\cite{resnet}), where all the different size variants have been tried. We used this architecture as it is a well established baseline for image classification and the size of the model is appropriate for the number of samples in the dataset. Pretrained \textit{Imagenet}~(\cite{imagenet}) weights from the \textit{torchvision} library (\cite{torchvision}) have been used to improve the training process. To adapt the pretrained model architecture to our task, first, we added zero padding to our one-channel image chunks to fit the 3-channel input (for RGB images), then, we removed the fully connected layer from the original model and we added our own classifier composed by: \textit{BatchNorm} and \textit{Linear(units=2048)} layers, followed by \textit{BatchNorm}, \textit{ReLU}, \textit{Dropout(p=0.4)} and the output \textit{Linear(units=17)} layer with \textit{Softmax} activation function.

As a result of this classification over all the chunks in an audio sample, we get a sequence of model predictions where we keep the probabilities for each class. Each prediction corresponds to the middle frame's timestep of each chunk. So we get a prediction every \textit{chunk\_hop} frames, which defines the resolution of our segmentation.

\subsection{ACS-DTW Segmentation Postprocessing}

The preliminary segmentation generated by the CNN model can contain some errors due to the fact that the model is not taking into account the order of the surfaces in the track. As this model is a classifier among all the surfaces but agnostic to the chronological order, given a chunk at time $m$ classified as belonging to the surface $n$, the model can classify the next chunk as any surface type, even the previous ones, when the unique valid surface labels are $n$ and $n+1$. Only these kind of operations are the possible ones due to the nature of the use case, where all the surfaces must appear in the correct order. Another problem generated by spontaneous misclassifications is that very short segments may appear, even with a duration of a single chunk. In this final step of the pipeline, we avoid these kind of errors in order to select the optimal surface boundaries. To accomplish this, we take into account the real sequence of surfaces and a minimum duration per surface.

The proposed method to generate the final segmentation is a dynamic programming algorithm with limited rules to (i) avoid assigning several labels to the same chunk while (ii) allowing that a subsequence of consecutive chunks are assigned to the same label. This is known as the Dynamic Time Warping (DTW). The original DTW compares two signals with different duration, $x$ and $y$, and returns an alignment between their elements based on a cost function which is usually minimized. In our case, we want to align the sequence of surfaces, $x$, in the track with the surface probabilities, $y$, predicted by the model, which is a sequence of predicted class probabilities for each chunk. The goal is to assign the final track label to each chunk by taking into account the real order of the surfaces. Consequently, the algorithm fills a cost matrix $D_{N \times M}$ by accumulating a cost function $c$, where $N$ is the number of surfaces and $M$ the number of chunks. And once $D$ is computed, the solution path can be extracted to get the alignment between the two sequences $x$ and $y$. In our case, this alignment assigns a surface label from $x$ to each chunk in $y$. Our cost function is the logarithm of the output probability of the CNN model for a given surface $x_n$ and chunk $y_m$: $c(x_n, y_m) = \log P(surface=n|chunk=m)$. Therefore, we maximize $c$ instead of minimizing it as the typical approach. In~\cite{dtw_review}, variations of the original DTW are presented to force the algorithm to take more suitable paths. Our proposed ACS-DTW is based on the variant to restrict the step function but with further modifications. These restrictions are applied to force the algorithm to set a minimum duration in chunks for each surface. Where the minimum duration per surface was extracted by computing the minimum duration of each surface in the training subset of the dataset. This duration in seconds is then converted to the corresponding number of chunks. The changes that we applied to our algorithm are the following:
\begin{itemize}
    \item \textbf{Remove vertical alignment:} In the original DTW three kinds of steps are allowed: (1) vertical, where an element of the sequence $y$ is aligned to more than one from $x$, (2) diagonal, where the elements of both sequences are aligned 1 to 1, and (3) horizontal, where an element of $x$ is aligned to more than one from $y$. Due to the problem restrictions, it does not make sense to assign a chunk from $y$ to more than one surface from $x$, so the vertical step is not considered in our implementation.
    \item \textbf{Diagonal step constraint:} As we removed the vertical step, the only way to advance to the next surface is by the diagonal move. Thus, instead of using the regular diagonal step, we applied a step function condition. This condition is that to be able to go to the next surface $n$, a minimum number of chunks $min\_chunks_n$ must be assigned to surface $n$, where $min\_chunks_n$ is the minimum number of chunks obtained from the dataset exploration for that surface $n$. Given this definition, the cost matrix $D$ is computed using the Equation~\ref{eq:cost_matrix}. Note that at each row $n$ the value of $min\_chunks_n$ will be different, adjusting the condition to each surface.
\end{itemize}

\begin{equation}
    \label{eq:cost_matrix}
    D(n, m) = \text{max} \begin{cases}
         D(n, m-1) + c(x_n, y_n), \\
         D(n-1, m-min\_chunks_n) + \sum_{i=1}^{min\_chunks_n} c(x_n, y_{n-i+1}),
    \end{cases}
\end{equation}

The full algorithm definition to fill the cost matrix is shown in Algorithm~\ref{algo:DTW}. Once we have the matrix $D$ computed, we can retrieve the alignment to get the final label for each chunk.

\begin{algorithm}[ht]
    \caption{\textbf{ACS Dynamic Time Warping (ACS-DTW):} Definition of the algorithm to fill the $D$ matrix. Note that the implementation of the backtrace to get the alignment is omitted for clarity.}
    \label{algo:DTW}
    \begin{algorithmic}[1]
        \renewcommand{\algorithmicrequire}{\textbf{Input:}}
        \renewcommand{\algorithmicensure}{\textbf{Output:}}
        \REQUIRE Surfaces sequence $x$, chunks predictions $y$, number of surfaces $N$, number of chunks $M$ and list of minimum chunks per surface $min\_chunks$
        \ENSURE  Cost matrix $D$
        \STATE // Initialization to maximize $c$
        \STATE $D(0..N, 0..M) = -\inf$
        \STATE
        \STATE // Fill the fist row of $D$ (surface $n=0$)
        \STATE $D(0, 0) = c(x_0, y_0)$
        \FOR {$m = 1$ to $M$}
            \STATE $D(0, m) = D(0, m-1) + c(x_0, y_m)$
        \ENDFOR
        \STATE
        \STATE // To skip a min number of chunks to reach a surface
        \STATE $start\_chunk = min\_chunks_0 - 1$
        \FOR {$n = 1$ to $N$}
            \STATE // Add the min number of chunks to reach surface $n$
            \STATE $start\_chunk = start\_chunk + min\_chunks_n$
            \FOR {$m = start\_chunk$ to $M$}
                \STATE \begin{equation*}
                    D(n, m) = \text{max} \begin{cases}
                         D(n, m-1) + c(x_n, y_n), \\
                         D(n-1, m-min\_chunks_n) + \sum_{i=1}^{min\_chunks_n} c(x_n, y_{n-i+1}),
                    \end{cases}
                \end{equation*}
            \ENDFOR
        \ENDFOR
        \RETURN $D$ 
    \end{algorithmic} 
\end{algorithm}

Finally, after getting the final chunks labels, we can transform that into timestamps (in seconds) to set the boundaries between surfaces.
\section{Experimental Setup} \label{sec:4_experimental_setup}

\subsection{Dataset}

The dataset used in this work has been collected at the Ford Manufacturing Plant in Valencia (Spain). All the audio data was recorded during production in order to get a representative dataset of the real environment.

To collect the samples a custom device was built to record audio. This device is designed to be placed in the headrest of the copilot and uses a Lavalier microphone to record mono audio. The configuration of the recording is set to use a sampling rate of 22050 Hz. And the data was stored in uncompressed WAV format to avoid any loss.

A data collection phase has been performed recording samples from the same vehicle model but with variations in terms of motorization and equipment. In total, 1823 audio samples were collected. Each of the recordings has an approximate duration of 2 minutes, including not only the audio from the track but also some additional audio before and after the track. These additional segments not belonging to the track are included because the driver starts the recording before entering the track and stops it outside the track. Nevertheless, this out-of-track segments count as a surface to detect. Because in case of using the proposed segmentation method inside a more complex pipeline, we may want to use it to remove these segments before doing any further audio analysis.

From the 1823 audio samples, we manually segmented 50 of them, by setting the exact instant of time where there is a transition between surface types. There are 17 different surfaces to detect, where two of them are corresponding to the preceding and succeeding segments not belonging to the track. As the track is completely standardized, it must be completed always at a determined speed and without stopping or changing the path. Therefore, we can take into account the constraint of always generating the sequence of surfaces in the same order and following a similar duration.

The 50 segmented audios have been selected to form a segmentation dataset, accumulating a total of 1.84 hours of audio. It was split in a train set of 42 samples and a test set of 8. Then during training, from the 42 samples, 8 were used for the validation set.

Because of the nature of the dataset the amount of data for each surface is not balanced, as the length and speed changes from one surface to other. For privacy reasons the exact distribution is not going to be shown, but to illustrate, the duration of a surface can go from 3 to 15 seconds approximately.

\subsection{Metrics}

The metrics used to evaluate the system can be divided in two groups:
\begin{itemize}
    \item \textbf{Classifier Metrics:} Metrics computed to evaluate the CNN model that classifies the individual chunks among the available track surfaces, without applying ACS-DTW algorithm. The classification metrics used are accuracy and F1-score.
    \item \textbf{Segmentation Metrics:} Metrics computed using the final segmentation after the segmentation postprocessing has been applied using our ACS-DTW algorithm. The metrics are:
    \begin{itemize}
        \item \textbf{Mean error:} Mean of the absolute error values when measuring distance between predicted and true surfaces boundaries.
        \item \textbf{Barrier Threshold Accuracy:} Accuracy metric applied to each one of the predicted boundaries. If the error (in seconds) for a boundary is higher than the threshold it is taken as a miss, else a hit. The thresholds considered are: 0.2s, 0.5s and 1.0s. These have been selected because they represent from a highly satisfactory threshold (0.2s) to the maximum (1.0s) that we consider acceptable for the use case.
    \end{itemize}
\end{itemize}

\subsection{Computing Resources} \label{subsec4:computing_resources}

The machine used to run the experiments has an Intel Xeon W-2245 CPU (8 cores), a NVIDIA RTX A6000 GPU (48 GB) and 32 GB of RAM. The batch size used is 512 as it results in the best performance.

\subsection{Software} \label{subsec4:software}

To build the pipeline several libraries have been used. Regarding data loading and preprocessing we used \textit{librosa}~(\cite{librosa}) and \textit{torchaudio}~(\cite{torchaudio}). To apply data augmentation to the waveform data we used \textit{audiomentations}~(\cite{audiomentations}). To orchestrate the training pipeline \textit{Pytorch Lightning}~(\cite{pytorch_lightning}) was used along with \textit{Optuna}~(\cite{optuna}) for hyperparameter optimization.
\section{Results} \label{sec:5_results}

In this section the results of our experimentation are shown. A comparison of different hyperparameters and training configurations is presented in order to analyze the impact of each variant on the CNN classifier and in the final segmentation.

To train all the experiments presented in the following tables, we used a baseline configuration with the \textit{AdamW} optimizer~(\cite{adamw}), a learning rate of $1\mathrm{e}^{-4}$, and a weight decay of $6.1\mathrm{e}^{-3}$. For feature extraction, we used the \textit{Mel spectrogram} with 70 Mel bands, and a \textit{chunk\_hop} of 1 spectrogram frame. As for the CNN classifier, we used the \textit{ResNet-152} CNN architecture with pretrained \textit{Imagenet} weights. This configuration was selected using \textit{Optuna} for hyperparameter optimization search.

Two of the most important hyperparameters to tune the performance of the system are: the size of the Fast Fourier Transform (FFT), declared as  \textit{n\_fft}; and the number of spectrogram frames per chunk, declared as \textit{chunk\_size}. Note that for the spectrogram computation using FFT, we use a window overlapping of 50\%. When tuning \textit{n\_fft} and \textit{chunk\_size}, the following trade-offs must be taken into account:
\begin{itemize}
    \item \textbf{\textit{n\_fft}}: Increasing the value decreases the total number of chunks extracted from the whole spectrogram, as the \textit{chunk\_hop} gets bigger in time. This implies to reduce the resolution of the segmentation but also improves the inference time, as less chunks have to be processed.
    \item \textbf{\textit{chunk\_size}}: The size of this value is directly proportional to the amount of context that we want to give per chunk. Small values may not provide enough context to effectively classify a chunk, but a very large context increases memory footprint and inference time.
\end{itemize}

Table~\ref{tab:fft_results} presents the results of our exploration of various combinations of \textit{n\_fft} and \textit{chunk\_size}. We selected different \textit{chunk\_size} values for each \textit{n\_fft} value to ensure that the amount of context included in each chunk was approximately the same in seconds. Our results indicate that classification metrics were lower for the smallest values of \textit{chunk\_size}, which resulted in decreased segmentation performance. After evaluating all the combinations, we determined that the combination of $\textit{n\_fft}=4096$ and $\textit{chunk\_size}=91$ was the optimal choice, striking a balance between accuracy and performance. While the combination of $\textit{n\_fft}=2048$ and $\textit{chunk\_size}=121$ resulted in a slight increase in precision, the cost was much higher, as it generates twice as many chunks and with higher size. The maximum chunk size in seconds is approximately 8.4, because in case of larger chunks, samples of the final class would not be generated due to the offset of the extraction window. Since there would be audios in which the duration of the last surface is less than half a chunk.

\begin{table}[htbp]
\large
\centering
\caption{Comparison of different sizes for the FFT computation (\textit{n\_fft}) in spectrogram transform and chunk\_size values. The \textit{chunk\_size} is adjusted for each value of \textit{n\_fft} to take a similar context window (in seconds).}
\label{tab:fft_results}
\resizebox{0.7\columnwidth}{!}{%
\begin{tabular}{@{}crccccccc@{}}
\cmidrule(l){6-9}
                         & \textbf{}                     & \textbf{}           & \textbf{}         & \textbf{}                & \multicolumn{4}{c}{\textbf{Barrier Threshold Acc.}} \\ \midrule
\textit{\textbf{n\_fft}} & \textit{\textbf{chunk\_size}} & \textbf{Chunk Acc.} & \textbf{Chunk F1} & \textbf{Mean err. (s)} & \textbf{0.2s}    & & \textbf{0.5s}    & \textbf{1.0s}    \\ \midrule
\multirow{6}{*}{1024}    & 81 (1.88s)                     & 0.76                 & 0.73               & 0.257                      & 0.75              & & 0.83              & 0.93                                          \\
                         & 161 (3.74s)                    & 0.83                 & 0.83               & 0.180                      & 0.79              & & 0.89              & 0.95              \\
                         & 241 (5.60s)                    & 0.85                 & 0.84               & 0.204                      & 0.75              & & 0.87              & 0.95              \\
                         & 281 (6.52s)                    & 0.82                 & 0.82               & 0.202                      & 0.78              & & 0.89              & 0.94              \\
                         & 321 (7.45s)                    & 0.85                 & 0.84               & 0.193                      & 0.79              & & 0.87              & 0.95              \\
                         & 361 (8.38s)                    & 0.87                 & 0.87               & 0.166                      & 0.76              & & 0.89              & 0.98              \\ \midrule
\multirow{6}{*}{2048}    & 41 (1.90s)                     & 0.73                 & 0.70               & 0.310                      & 0.67              & & 0.83              & 0.91              \\
                         & 81 (3.76s)                    & 0.85                 & 0.84               & 0.197                      & 0.76              & & 0.90              & 0.96              \\
                         & 121 (5.62s)                    & 0.86                 & 0.85               & \textbf{0.157}                      & \textbf{0.82}              & & 0.91              & 0.97              \\
                         & 141 (6.55s)                    & 0.85                 & 0.83               & 0.271                      & 0.76              & & 0.87              & 0.92              \\
                         & 161 (7.48s)                    & 0.87                 & 0.86               & 0.299                      & 0.78              & & 0.88              & 0.95              \\
                         & 181 (8.41s)                    & 0.88                 & 0.86               & 0.181                      & 0.81              & & 0.90              & 0.98              \\ \midrule
\multirow{6}{*}{4096}    & 21 (1.95s)                     & 0.72                 & 0.67               & 0.261                      & 0.68              & & 0.85              & 0.96              \\
                         & 41 (3.81s)                     & 0.83                 & 0.81               & 0.303                      & 0.69              & & 0.88              & 0.94              \\
                         & 61 (5.67s)                     & 0.85                 & 0.83               & 0.230                      & 0.68              & & 0.90              & 0.96              \\
                         & 71 (6.59s)                     & 0.88                 & 0.86               & 0.226                      & 0.73              & & 0.92              & 0.96              \\
                         & 81 (7.52s)                     & 0.89                 & 0.88               & 0.180                      & 0.76              & & \textbf{0.93}              & 0.98              \\
                         & 91 (8.45s)                    & \textbf{0.91}                 & \textbf{0.90}               & 0.178                      & 0.77              & & 0.90              & \textbf{0.99}              \\ \bottomrule
\multirow{6}{*}{8192}    & 11 (2.04s)                     & 0.54                 & 0.53               & 0.480                      & 0.53              & & 0.74              & 0.86              \\
                         & 21 (3.90s)                     & 0.71                 & 0.67               & 0.423                      & 0.55              & & 0.77              & 0.88              \\
                         & 31 (5.76s)                     & 0.83                 & 0.81               & 0.362                      & 0.61              & & 0.86              & 0.91              \\
                         & 35 (6.50s)                     & 0.82                 & 0.82               & 0.376                      & 0.62              & & 0.85              & 0.94              \\
                         & 41 (7.61s)                     & 0.86                 & 0.86               & 0.408                      & 0.64              & & 0.84              & 0.94              \\
                         & 45 (8.36s)                    & 0.87                 & 0.86               & 0.240                      & 0.65              & & 0.90              & 0.97              \\ \bottomrule
\end{tabular}%
}
\end{table}

After analyzing the chunk extraction parameters, we determined that the optimal combination is $\textit{n\_fft}=4096$ and $\textit{chunk\_size}=91$, from this baseline, we further investigated how different sizes of our model could affect performance. In previous experiments, we used the largest variant, \textit{ResNet-152}. In Table~\ref{tab:resnet_results}, we present the results of our experiments using smaller versions of \textit{ResNet}. Interestingly, the best results were obtained using the smallest variant, \textit{ResNet-18}. This suggests that using a model with fewer parameters can improve performance without sacrificing accuracy.

\begin{table}[htbp]
\large
\centering
\caption{Comparison of the different ResNet model variants with pretrained weights. All the experiments where run with the same data preprocessing and training configuration ($\textit{n\_fft}=4096$ and $\textit{chunk\_size}=91$).}
\label{tab:resnet_results}
\resizebox{0.7\columnwidth}{!}{%
\begin{tabular}{@{}cccccccc@{}}
\cmidrule(l){5-8}
\textbf{}       & \textbf{}               & \textbf{}         & \textbf{}                & \multicolumn{4}{c}{\textbf{Barrier Threshold Acc.}}                       \\ \midrule
\textbf{ResNet} & \textbf{Chunk Accuracy} & \textbf{Chunk F1} & \textbf{Mean err. (s)} & \textbf{0.2s} & & \textbf{0.5s} & \textbf{1.0s} \\ \midrule
18              & \textbf{0.92}            & \textbf{0.92}      & \textbf{0.174}             & 0.73            & & \textbf{0.90}  & \textbf{1.0}   \\
34              & \textbf{0.92}            & 0.91               & 0.202                      & 0.73            & & \textbf{0.90}  & 0.96           \\
50              & 0.85                     & 0.84               & 0.347                      & 0.63            & & 0.83           & 0.94           \\
101             & 0.89                     & 0.88               & 0.257                      & 0.70            & & 0.81           & 0.97           \\
152             & 0.91                     & 0.90               & 0.178                      & \textbf{0.77}   & & \textbf{0.90}  & 0.99           \\ \bottomrule
\end{tabular}%
}
\end{table}

We explored the \textit{ResNet} architectures using pretrained weights from \textit{Imagenet} as a baseline. It has been demonstrated that transfer learning using \textit{Imagenet} weights can improve results even in different domains~(\cite{imagenet_tranfer_learning}). However, we also compared the results using the same \textit{ResNet-18} architecture trained from scratch and found that using pretrained weights consistently improved the training results (Table~\ref{tab:pretrained_results}).

\begin{table}[htbp]
\large
\centering
\caption{Comparison of using a \textit{ResNet-18} with pretrained weights and without.}
\label{tab:pretrained_results}
\resizebox{0.7\columnwidth}{!}{%
\begin{tabular}{@{}cccccccc@{}}
\cmidrule(l){5-8}
\textbf{}                   & \textbf{}               & \textbf{}         & \textbf{}                & \multicolumn{4}{c}{\textbf{Barrier Threshold Acc.}}                       \\ \midrule
\textbf{Pretrained Weights} & \textbf{Chunk Accuracy} & \textbf{Chunk F1} & \textbf{Mean err. (s)} & \textbf{0.2s} & & \textbf{0.5s} & \textbf{1.0s} \\ \midrule
Yes (Imagenet)              & \textbf{0.92}           & \textbf{0.92}     & \textbf{0.174}         & \textbf{0.73} & & 0.90          & \textbf{1.0}  \\
No                          & 0.90                    & 0.91              & 0.188                  & 0.70          & & \textbf{0.93} & 0.97          \\ \bottomrule
\end{tabular}%
}
\end{table}

In addition to the Mel spectrogram, which we used to extract frequency-domain features, we explored two other transformations: the \textit{Base} spectrogram, which was extracted using the Short Time Fourier Transform (STFT), and Mel-Frequency Cepstrum Coefficients (\textit{MFCC}), which were extracted using 40 coefficients. Table~\ref{tab:spectrograms_results} presents the results of our experiments with these different spectrograms. We found that the \textit{MFCC} method resulted in significantly worse classification metrics for the chunks and a high segmentation error, indicating that it is not suitable for this task. As for the other two methods, \textit{Base} and \textit{Mel}, we obtained similar results in terms of segmentation accuracy with no clear best option. However, considering inference time and memory footprint, the Mel spectrogram is a better option, as it generates chunks that are much smaller.

\begin{table}[htbp]
\large
\centering
\caption{Comparison of different transforms to extract a spectrogram from the waveform data. Where \textit{Base} represents the Short Time Fourier Transform (STFT) spectrogram, \textit{Mel} the Mel spectrogram using 70 Mel bands, and \textit{MFCC} the Mel-Frequency Cepstrum Coefficients with 40 coefficients.}
\label{tab:spectrograms_results}
\resizebox{0.7\columnwidth}{!}{%
\begin{tabular}{cccccccc}
\cmidrule(l){5-8}
\textbf{}            & \textbf{}               & \textbf{}         & \textbf{}                & \multicolumn{4}{c}{\textbf{Barrier Threshold Acc.}}                       \\ \midrule
\textbf{Spectrogram} & \textbf{Chunk Accuracy} & \textbf{Chunk F1} & \textbf{Mean err. (s)}  & \textbf{0.2s} & & \textbf{0.5s} & \textbf{1.0s} \\ \midrule
Base                 & \textbf{0.92}                     & \textbf{0.92}               & \textbf{0.172}                      & \textbf{0.73}            & & \textbf{0.92}           & 0.99           \\
Mel                  & \textbf{0.92}                     & \textbf{0.92}               & 0.174                      & \textbf{0.73}            & & 0.90            & \textbf{1.00}           \\
MFCC                 & 0.56                    & 0.57               & 1.945                      & 0.41            & & 0.60            & 0.71           \\ \bottomrule
\end{tabular}%
}
\end{table}

As our dataset contains only $1.84$ hours of audio, we explored different data augmentation methods to improve the training results. We experimented with waveform level transformations and \textit{MixUp} applied to the training batches as a generic data augmentation method. Table~\ref{tab:DA_results} presents the results of each of these methods and their combination. We found that the accuracy metrics decreased when waveform augmentation was applied, indicating that it is not suitable for this task. However, in the case of \textit{MixUp}, we observed that while the CNN model accuracy was lower, the segmentation precision was higher. This suggests that the probabilities given by the model are better calibrated when using \textit{MixUp}, and as a result, the ACS-DTW algorithm can achieve a slightly more precise segmentation.

\begin{table}[htbp]
\large
\centering
\caption{Comparison of different Data Augmentation methods. The methods compared are: (1) No data augmentation, (2) Waveform transforms, (3) Batch MixUp, (4) Combination of waveform transforms and batch MixUp.}
\label{tab:DA_results}
\resizebox{0.7\columnwidth}{!}{%
\begin{tabular}{@{}cccccccc@{}}
\cmidrule(l){5-8}
\textbf{}                        & \textbf{}               & \textbf{}         & \textbf{}                & \multicolumn{4}{c}{\textbf{Barrier Threshold Acc.}}                       \\ \midrule
\textbf{Data Aug.}               & \textbf{Chunk Accuracy} & \textbf{Chunk F1} & \textbf{Mean err. (s)} & \textbf{0.2s} & & \textbf{0.5s} & \textbf{1.0s} \\ \midrule
No                  & \textbf{0.92}                     & \textbf{0.92}               & 0.174                      & 0.73            & & 0.90            & \textbf{1.0}           \\
Wave                             & 0.90                     & 0.89               & 0.198            & 0.66            & & 0.87           & 0.99           \\
MixUp                            & 0.89                     & 0.89               & \textbf{0.166}            & \textbf{0.79}            & & \textbf{0.92}           & 0.98           \\
\multicolumn{1}{l}{Wave + MixUp} & 0.88                     & 0.89               & 0.281            & 0.71            & & 0.90           & 0.97           \\ \bottomrule
\end{tabular}%
}
\end{table}
\section{Conclusions} \label{sec:6_conclusions}

In this work, we presented the ConvDTW-ACS method for accurately segmenting track surface types in audio recordings of vehicles driven through a production test track. Our method was evaluated on a real-world dataset collected from the Ford Manufacturing Plant in Valencia (Spain), achieving a mean error of $166$ milliseconds when delimiting the boundaries of the track surfaces. These results demonstrate the effectiveness of the proposed method in accurately segmenting different surface types, enabling the application of more specialized AI systems such as Anomaly Detection to improve the quality inspection process. Moreover, we explored various hyperparameter combinations and presented their trade-offs in terms of segmentation accuracy and computational cost. Overall, our proposed method provides a promising approach for improving the efficiency and accuracy of quality inspection in vehicle production lines.
\section{Acknowledgments} \label{sec:7_acknowledgments}

We would like to thank the Ford Manufacturing Plant in Valencia (Spain) for supporting this project and making this research possible. Especially, we would like to thank the Quality Department for their great involvement in the correct collection and labelling of the data; and the Launch Department for promoting and supporting the project.

This research was funded by Ford Motor Company under the Ford Manufacturing Plan in Valencia (Spain) and the KU Leuven Ford Alliance Agreement (Project: Automated Squeak \& Rattle Testing, KUL0134).

\printbibliography

\end{document}